July 15, 2020

# Senior Living Communities: Made Safer by AI

**Ashutosh Saxena** and **David Cheriton**

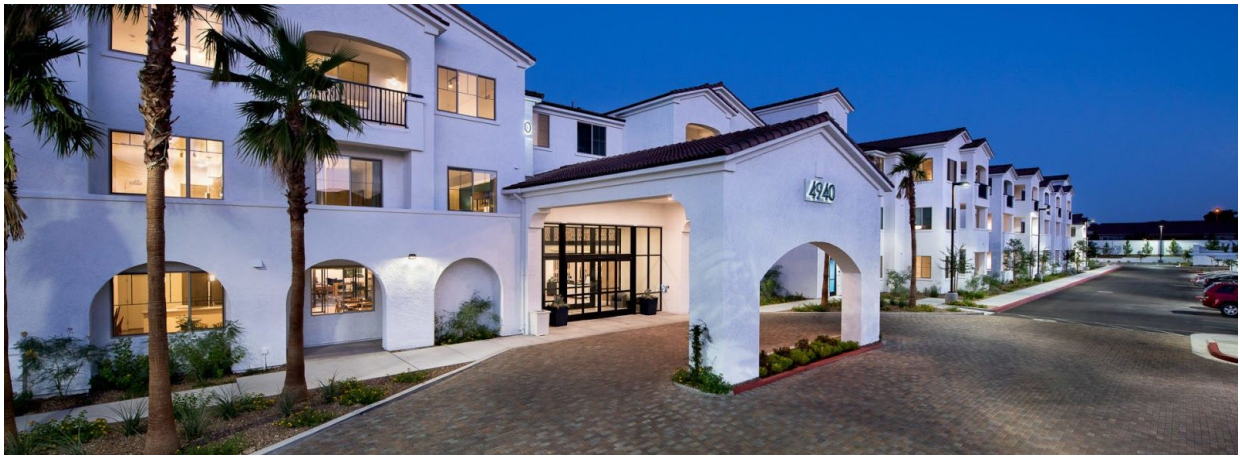

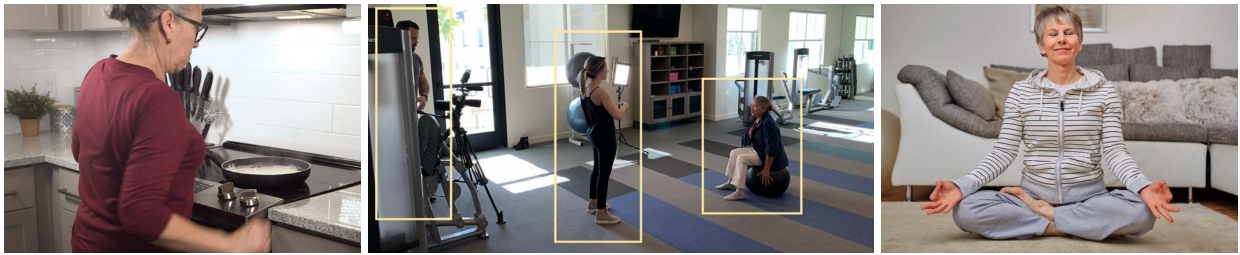

**Ashutosh Saxena** is CEO and co-founder of Caspar.AI. He received his PhD in Computer Science from Stanford University with advisor Andrew Ng. He has been recognized as one of Eight Innovators to Watch by Smithsonian Institution, TR35 Innovator by MIT Technology Review, and a Sloan Fellow.

**David Cheriton** is a Silicon Valley investor who co-founded Arista Networks (IPO), Kealia (Acq. Sun Microsystems), and Granite Systems (Acq. Cisco Systems). He wrote Google their first check and is listed on the Forbes Midas List. He is also a Professor of Computer Science at Stanford University, and received the SIGCOMM Lifetime Achievement Award.

# Senior Living Communities: Made Safer by AI

There is a historically unprecedented shift in demographics towards seniors, which will result in significant housing development over the coming decade. This is an enormous opportunity for real-estate operators to innovate and address the demand in this growing market.

However, investments in this area are fraught with risk. Seniors often have more health issues, and Covid-19 has exposed just how vulnerable they are – especially those living in close proximity. Conventionally, most services for seniors are "high-touch", requiring close physical contact with trained caregivers. Not only are trained caregivers short in supply, but the pandemic has made it evident that conventional high-touch approaches to senior care are high-cost and greater risk. There are not enough caregivers to meet the needs of this emerging demographic, and even fewer who want to undertake the additional training and risk of working in a senior facility, especially given the current pandemic.

In this article, we rethink the design of senior living facilities to mitigate the risks and costs using automation. With AI-enabled pervasive automation, we claim there is an opportunity, if not an urgency, to go from high-touch to almost "no touch" while dramatically reducing risk and cost.  Although our vision goes beyond the current reality, we cite measurements from Caspar AI-enabled senior properties  that show the potential benefit of this approach.

# Contents

# 1 Introduction

There is a historically unprecedented shift in demographics towards seniors – about 30% of the population in the United States is 55 years or older, expected to double by 2050 [5] (Figure 1). This will result in significant housing development over the coming decade [8]. *"As the industry envisions the future of design, building, and operations of senior housing there is a great opportunity to optimize the health, safety, and living experience of boomers."* said Fritz Wolff, of The Wolff Company, ranked #12 by NMHC in 2019 as 25 Largest Apartment Developers in the US [34].

However the investments in this area are fraught with risk (e.g., operationally intensive [15], prone to litigation [9], and high insurance rates [28]). Covid-19 has exposed just how vulnerable seniors living in senior living facilities are [7]. A lot of Covid-related sickness and deaths have occurred in high-density senior housing [36] – e.g., one Mardi Gras party at a Kirkland senior living community led to 35 deaths [6]. The risk of a pandemic has been known for years [4] and Covid-19 has made that risk a reality. Covid-19 is not the only threat that seniors face [4] – with their compromised immune systems many in senior living facilities die of other contagious diseases every year. Additionally, time-critical events, for instance falling from a stroke, are often detected too late [13].

Adding to the risk, the government and CDC are changing regulations on how you need to operate the facilities [10]. Conventionally, most services require *physical proximity of trained caregivers* [7]. Not only are trained "high-touch" caregivers in short supply [14] and only available at an increased cost (Figure 2), the associated risks exacerbate the problem (e.g., [6,9]). Even providing basic services like maintenance and organizing onsite events are now an operational risk.

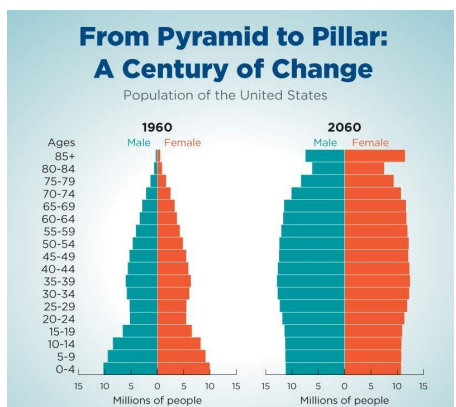
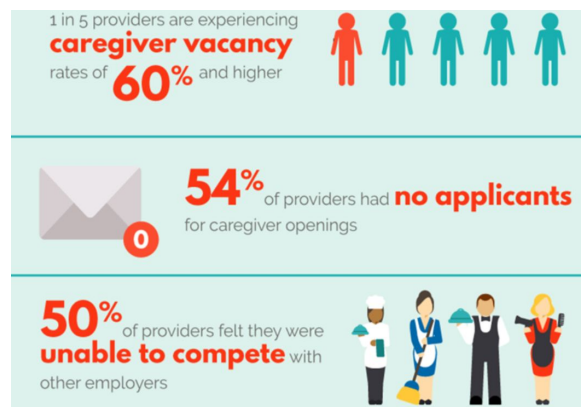

*Figure 1. There is an unprecedented demographic shift for seniors. About 30% of people in the US are 55 years or older, expected to double by 2050 [5].*

*Figure 2. "Senior communities face an acute shortage of trained caregivers. The timing is right for tech innovation." said Laurie Orlov, Analyst at Aging and Health Technology Watch.*

In this article, we rethink the design of senior living facilities to mitigate the risks and costs discussed earlier. With AI-enabled services, there is an opportunity to do things differently – you can design the buildings for *pervasive automation*. We define the problems in Section 2, present pervasive automation as a solution in Section 3, and present the case studies in Section 4 of a senior facility with significant automation provided by Caspar.AI, a smart home solutions company.

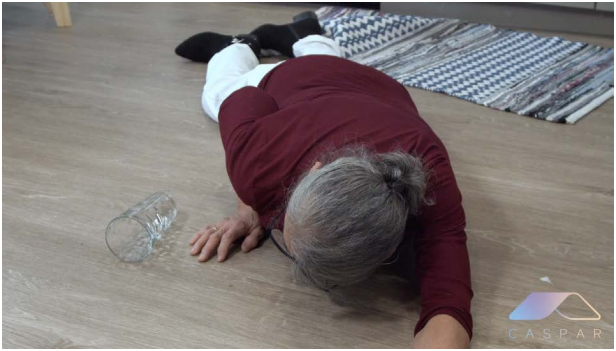

*"My biggest worry is that something would happen to her and no one would even know."*

*Figure 3. Safety of the residents is the primary concern when moving into a senior living facility. A high-touch approach may now discourage prospective renters from moving into a community because of perceived risks.*

# 2 Conventional Senior Living Needs Rethinking

As a senior housing operator, your goal is to attract, retain and delight the residents. Often the senior resident and their family make joint decisions. Before the prospective residents move into the property, they and their family will likely ask a variety of questions.

## 2.1 Will my mother be safe in your apartment?

The prospective resident's family says, "*My biggest worry is that something would happen to her and no one would even know.*" The families worry about their elderly loved ones living alone, tripping and falling (Figure 3), forgetting to turn the stove off (Figure 4), getting ill, or simply being lonely.

You explain to her about the onsite caregiver staff who do daily check-ins – they go unit by unit to check the well-being of the resident. In the past, they would have been happy with this answer. Now they ask the pandemic questions, *"Who comes in?" "How do you ensure safety? "Do they wear masks?" "How do you ensure they are not bringing virus in the apartment?"* To them, having unknown visitors already carries a negative stigma [6,7]. Now with Covid-19, the potential risk of seniors with fragile immune systems living in close proximity to each other is at the forefront of their mind [4,13]. This adds additional burden on your property – planning for masks, disinfecting, training, and even then there are still legal risks from *both* caregivers and residents.

In summary, in the past the high-touch approach was a part of the solution. Now it is a problem.

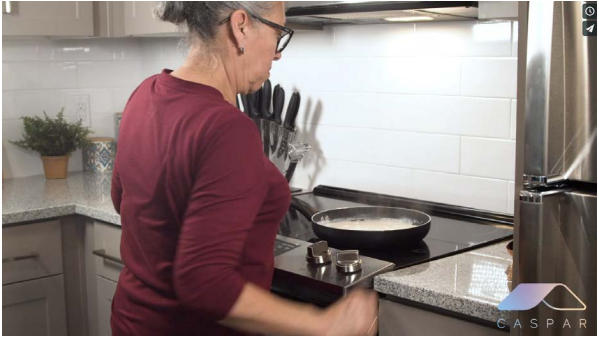 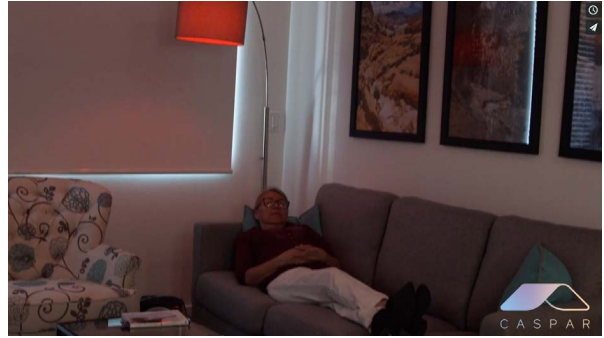

*A resident is cooking in the kitchen.*   *The stove is left on, she forgets and sleeps.*

*Figure 4. Forgetfulness can cause safety issues for the resident as well as the property.*

## 2.2 How do you safely maintain the apartments and facilities?

Prospective senior residents ask, *"How do you schedule visits? How does she (the resident) tell you that something is broken? Do you do proactive maintenance?"* Your staff does a variety of tasks, however it is challenging to do them automatically and proactively. This is especially important for many seniors who may not be proactive themselves about requesting maintenance. For example, a senior resident may not regularly look under the sink, so a water leak could go unreported for days.

Now they ask the pandemic questions, *"But how do you minimize the spread of infectious diseases?"* Residents do not want unnecessary maintenance visits, site staff, or non-essential personnel to enter their living space. The pandemic has changed how they want to interact with the outside world.

They ask, *"Do you track visitors? Do you take their temperature?"* You explain that your staff uses hand sanitizer before they enter the apartment, following government regulations [29]. But they remain worried and ask, *"But what if they visited an infected resident before coming to my mother's apartment? Can they schedule the visit when she is not at home?"*

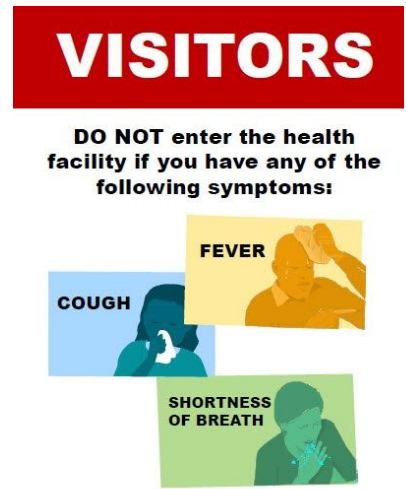

*Figure 5. Guidelines by CDC [29] are becoming an expectation by the residents, especially in case of an incident.*

These questions and concerns make it challenging for you to maintain high-quality service. To mitigate the risks, there is now an additional burden imposed on you to make an operational playbook (Figure 5), spend additional resources to train your staff, and take additional precautions to minimize the risk of an outbreak. We discuss how our pervasive automation addresses these problems in Section 3.2.

## 2.3 Will my mother stay active and have social contact?

The family says, "*I am worried about my mother's mental health, especially with social distancing. How do you engage residents socially?"* The families care about the seniors having an engaged lifestyle as it leads to motivation [18,19] – an active and engaged lifestyle reduces loneliness and functional decline [23, 24, 25, 30]. You show them the different activity rooms you have on the property for an enhanced social life (Figure 6) and mention the group classes you offer like Yoga for a healthy lifestyle.

Now they ask the pandemic questions, "*Do you check if the Yoga teacher is infected? How do you ensure that an infected resident is not attending?"* You had designed the common areas (Figure 6) with good intentions, but now the residents see it as a cause for concern. In a community living facility, one person being sick with an infectious disease can spread throughout the facility. In fact, at a Kirkland facility, there was a major outbreak of Covid-19 because of a Mardi-Gras party in a shared social space [6].

They further ask , "*Do you practice social distancing? How do you check that common areas aren't too crowded?"* They now expect you to regulate the use of Common areas for safety considerations [10]: follow social distancing and enforce density rules. This increases the burden of enforcement on your staff. Now it is a huge liability to run social programs with high-touch caregivers.

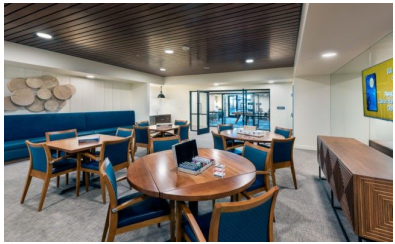   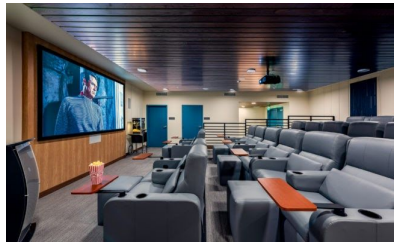   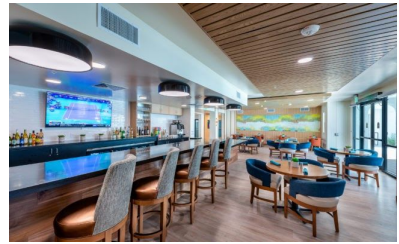

*Game room*  *Theater room*  *Party room*

*Figure 6. Common areas were designed with good intentions in mind. Now they are a liability.*

Overall, in this post-pandemic world, the conventional approach of high-touch caregiving which used to be a big part of the solution now has major problems:
1. *Resident Expectations.* Residents may now interpret a well-intentioned attempt of a caregiver to engage socially to help onsite as not only privacy-intrusive but also an unnecessary risk.
2. *Regulatory hurdles and legal risks.* Not just training caregivers, you now face a challenge of enforcing regulations on how your staff and visitors go on property to avoid potential legal issues from employees or from residents [6,7,10].
3. *Availability of trained, care-giving staff.* You face a challenge to get well-trained caregiving staff. Their availability, quality and cost is one of the biggest hurdles to cross [14].

We think pervasive automation can be a solution, which we present in the next section.

# 3 Pervasively Automated Safe Senior Living (PASSL)

Senior living can be made dramatically safer with pervasive automation. This automation also significantly reduces the costs and risks for operators, enabling you to go from *high-touch to (almost) no-touch.*

Pervasive automation is a system that uses the sensors to understand what is happening in the building, and controls functionalities such as climate, lighting and locks, and sends notifications to operators. In this section, we will illustrate how pervasive automation addresses the problems described in Section 2.

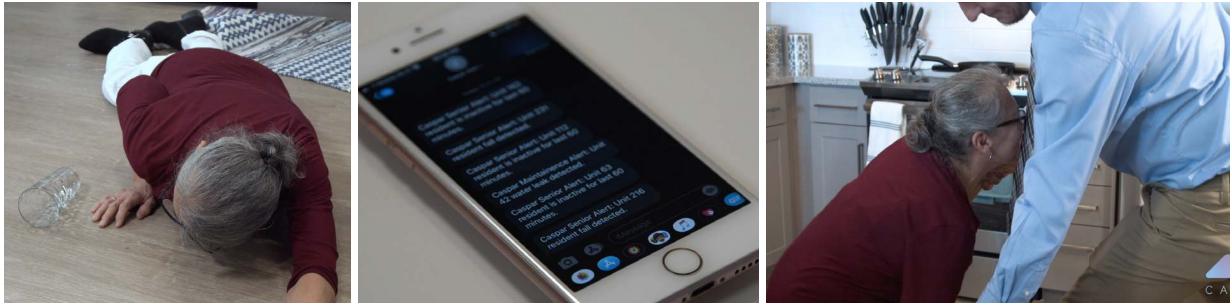

*Figure 7. With pervasive automation, the home can detect incidents such as inactivity or a fall. An alert is sent to onsite staff to help.*

## 3.1 Safe Living in Apartments with Automation

With pervasive automation, we can detect incidents such as resident inactivity, the stove left on or a fall – which are big concerns for the resident's family (Section 2.1). For example, if the sensors detect that the resident fell (Figure 7) or stove was left on (Figure 9), the system knows it is a safety risk.

The system can then alert the onsite staff. Such automated alerts run 24/7. In some cases, such as a stroke, alerting about the incident in a timely manner can be life saving. Since the staff only needs to go to the unit when there is an incident, it also reduces the risk of infection. Our system can also report other anomalies such as those related to daily movement levels, sleep quality, coughs, water intake, heart-rate, temperature, fall risk from gait change, and more – this could be useful to detect if someone is ill in the building.

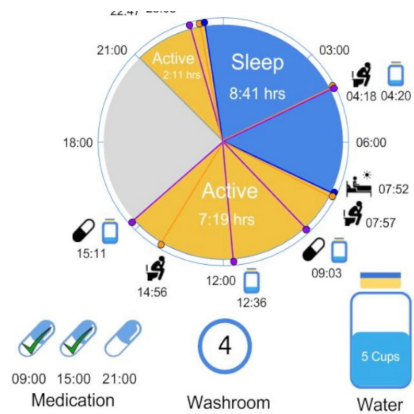

In contrast, conventional check-ins require the staff going in physically, typically once every 24 hours. The staff has to then send manual notifications to the resident's family.

*Figure 8. Analytics about a person's lifestyle supports various applications.*

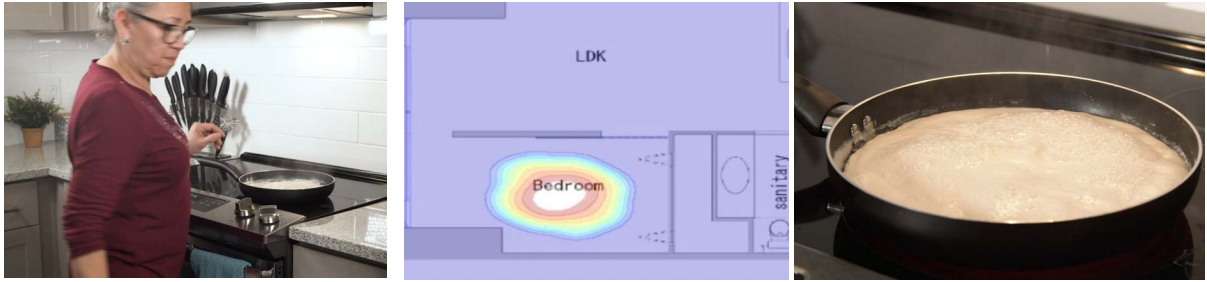

| Resident turns the stove on to cook. | Sensors detect that she went to the bedroom. | Stove was left on, automation turns it off for safety. |

*Figure 9. For safety alerts, our AI system uses sensors to obtain resident location and activities, and then takes actions to keep the residents safe.*

Healthcare partners can use data provided by pervasive automation to monitor the medical symptoms of residents automatically. Sensor data (Figure 8) could be useful in early detection of illnesses such as dementia [17, 21] and Parkinson's [22]. With the growing acceptance of telehealth [35], hospital visits for simple diagnoses can be avoided with pervasive automation, hence reducing the risk of infection. Of course, this system can be turned off if a resident is concerned more about their privacy than safety. Furthermore, we can restrict which information goes to which party – resident vs operator vs telehealth.

An automated home can do all the aforementioned items without requiring the resident to wear any device. Wearable devices (e.g., FitBit or necklace) can be useful extensions to our smart building system. They give activity reports and can detect falls. However, seniors often do not wear the device, either because of a stigma or because they feel uncomfortable using it (they take it off at night or in the bathroom). In a pervasively automated home, no devices are visible except a wall panel (Figure 10).

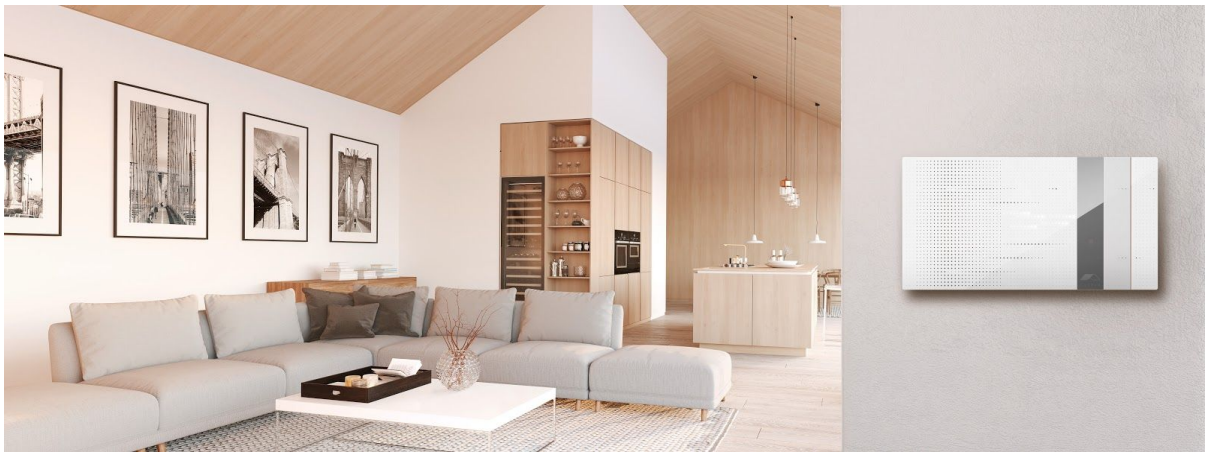

*Figure 10. In this pervasively automated apartment, no devices are visible except a wall panel.*

Pervasive automation can enable numerous amenities for engaged living – a consideration that resident's family have when moving their senior into the community. The automated home can motivate the residents to lead a healthy life, with more exercise, high-quality sleep, healthy food, and sufficient water intake [13]. As an example, the home adapts the environment of the home – during sleep time slowly dim the lights for a soothing ambiance. When the resident goes to the bathroom at night, dim lights turn on to help the person walk safely. The home can open shades automatically in the morning to let in natural light for motivation [18].

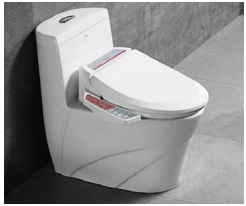 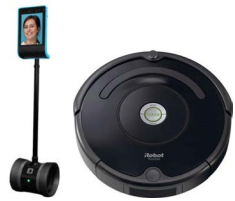 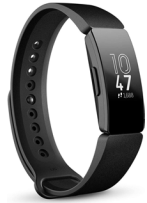 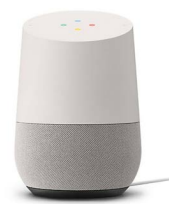

*Smart toilets*  *Telepresence and cleaning robots*  *Wearable widgets*  *Smart speakers*

*Figure 11. Pervasive automation can add numerous amenities for safe and convenient living.*

The system can also manage additional amenities such as smart toilets, cleaning robots and telepresence (Figure 11). Box solutions such as Alexa or Google Home can also provide additional functionality to the building system, but they are difficult to set up, customize, and maintain. This increases the work of the onsite staff, instead of reducing it. Smart functionality should be a seamless part of the building as an integrated system, just like air-conditioning – a central system that is easy to manage and operate.

## 3.2 Maintenance and Service with Automated Analytics

With pervasive automation, maintenance problems are automatically detected and alerts are created. For example if there is a water leak, then timely detection can avoid damage. Thus, routine checks can be replaced with checks only on an as-needed basis.

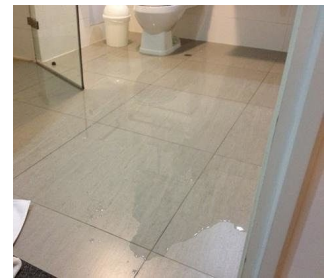

Pervasive automation can help schedule these visits when the senior is not around. This reduces the infection risk. For example, if the apartment is showing an anomaly in the air-conditioning running when it should not, it could be detected. Furthermore, this air-conditioning service can be scheduled when the resident goes out to breakfast.

*Tap is left on with water overflowing. An alert is automatically created.*

The key concerns of the residents in Section 2.2 are about proactive maintenance and scheduling service visits to minimize the contact risk. Safety and health of staff working in senior living communities also increases the risk. With automated maintenance alerts and scheduling, the infection risk can be substantially reduced in regular tasks such as cleaning, maintenance or package delivery.

With automation, you can improve maintenance efficiency, provide more services for residents, and go from a high-touch approach to a low-touch one.

## 3.3 Safe use of Common Areas with Automation

With pervasive automation, sensors in the common areas monitor who is there, their activities and if the number of people exceeds a predetermined limit.

As mentioned in Section 2.2, government regulations require measures to be taken for density management. In common areas such as gym, game room, movie room, corridors, elevators and model units, pervasive automation can use sensing to detect the number of people in the building. It can also help in maintaining social distancing by raising alerts in case the number of people is exceeded or someone who is not part of a scheduled group is there.

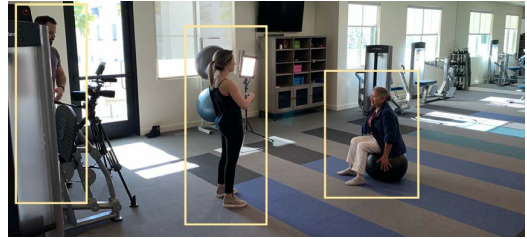

*Figure 12. Automation can estimate the number of people in an area, and suggest schedules to help in social distancing.*

To help further, schedules can be recommended to the residents about availability of the common areas. For example at breakfast time, it can help scheduling visits and letting residents know when the number of people is lower in the common areas. Thermal cameras can detect the temperature, and pervasive automation can perform contact tracing by providing a trackable and auditable event history.

In addition, pervasive automation enables the use of common areas in a touch-free and remote manner. The lighting, shades and climate can be set automatically to the preferred ambiance. Conventionally, it takes about 10 minutes for staff to go to each common area to change the settings, for which they need to touch multiple devices by hand. Automation reduces this work as well as decreases the risk of infection.

There are numerous other applications that are enabled by pervasive automation, such as smart access in the building, analytics about visitors – known vs unknown, and self-guided tours (Section 4). Automation also saves energy and provides analytics to the operators. With such business intelligence, operators can provide safe and engaged living for residents while reducing operational costs.

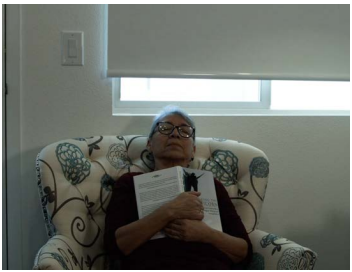
*Home wakes up the resident with natural lighting*

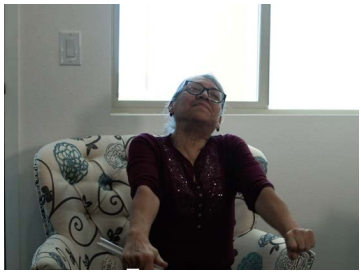
*Suggestion to go to the gym is made when it is available.*

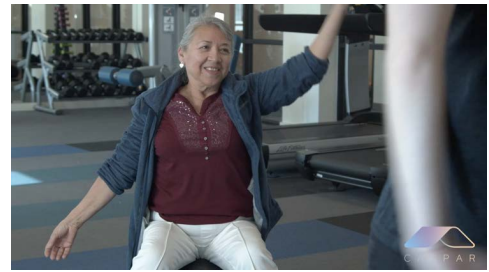
*The resident is able to use the gym with and still maintain social distance.*

*Figure 13. With pervasive automation, the common areas can be enabled for effective use – with seamless experience for residents from their home to the common areas like gym.*

# 4 Case Study: Automated Senior Living

We have implemented automation at several senior communities over the past three years, reducing high-tough caregiving to low-touch. In this section, we use two properties as case studies – Revel Nevada (Las Vegas, NV) and Revel Rancharrah (Reno, NV) – approximately 300 apartments.

*"In 2017, we engaged Caspar.AI to install their automated building system to increase the safety, comfort, and convenience for our senior residents"* said Fritz Wolff, of The Wolff Company.

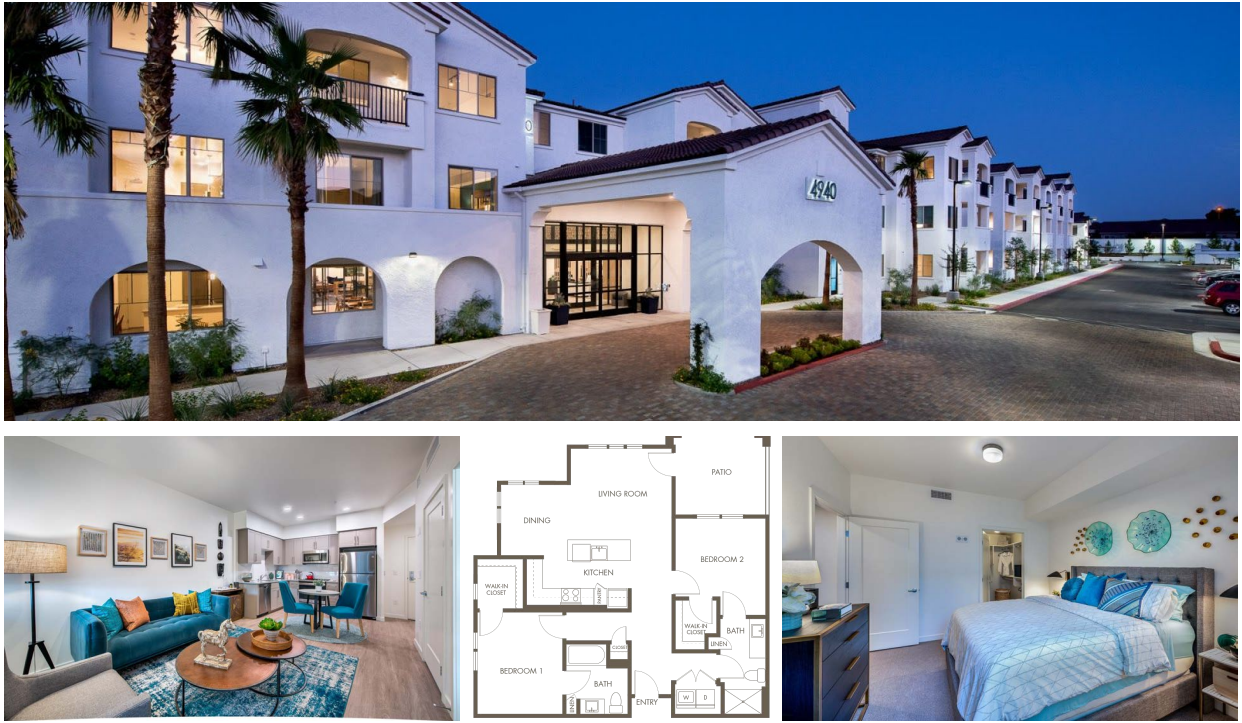

*Figure 14. Revel in Las Vegas, NV – a senior community built-in 2018. It promotes a healthy lifestyle with best-in-class wellness services and community amenities. Caspar AI-enabled pervasive automation in apartments and common areas provides a variety of services for increasing the safety of the residents and caregiving staff.*

We equipped all the apartments and common areas with smart lighting, fans, thermostat, motorized shades, motion sensors, speakers, microphones, bluetooth, wifi, and outdoor cameras. Our system is installed with simple, drop-in replacement devices. This results in about 10,000 devices that our system uses to provide a wide-variety of services to the residents. Some residents choose to add more devices such as a smart coffee machine, cleaning robot, and pet feeder.

Our AI-enabled automation assists the residents in their apartment, while providing them with numerous safety features. It also enables the staff to manage the building and provide services remotely.

## 4.1 Safe and Engaged Living in Revel Apartments

As discussed in Section 2.1, the resident's family is most worried that something would happen to their mother without anyone knowing. We address their concerns with AI-based notifications that send updates about the resident's most recent interaction with our system and movement in the apartment. Therefore, there is no need for the staff to enter the apartment to check if she is doing fine and risk infection – this feature allows the staff to go from a high-touch solution to a no-touch solution.

Residents' families receive updates 24/7 rather than just once a day. This solution is built into the apartment, and does not require the resident to wear any device. We estimate that the property staff reduces the time spent in checking in by about 28% per day by prioritizing the physical visits (Section 5).

Furthermore, this also helps the staff because the sensor-based status, and notes from the in-person visits are available to the resident's family immediately. This reduces the time the staff has to spend relaying information about the resident to the residents' families.

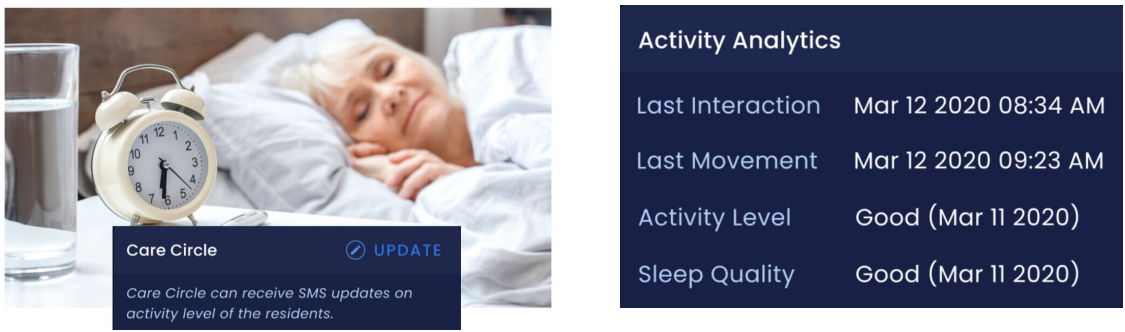

> *"I can finally sleep well knowing that my mother is safe and well cared for, especially in these challenging times."*
>
> - *family member of a senior resident at Revel.*

*Figure 15. If a senior becomes inactive, their family knows about it. Families love this as it gives them peace of mind 24/7. For the property staff, it is ideal as they do not risk infection by physically checking on the residents.*

In addition, automation encourages an active lifestyle with more exercise, high quality sleep, healthy eating habits, and sufficient water intake [18,19,30,23,24,25]. Caspar AI helps by providing an automated lifestyle coach that helps residents lead healthier lives. Our activity tracker informs the residents about how active they are (Figure 15, left) – a study found that activity tracking is seen as a helpful motivational tool in making lifestyle changes [31].

To increase motivation and create an environment of engaged living, Caspar AI changes the home environment itself – automatically modifying climate, lighting, and shades without requiring physical intervention from the resident. For example, at night the sleep scene sets the ambiance to help improve the sleep quality and provide automatic night lights for additional safety. In the morning Caspar AI plays soothing music and opens the shades to let in sunlight (Figure 17), which has been shown to increase motivation [18]. It can even set up a Yoga scene to encourage exercise (Figure 16 right).

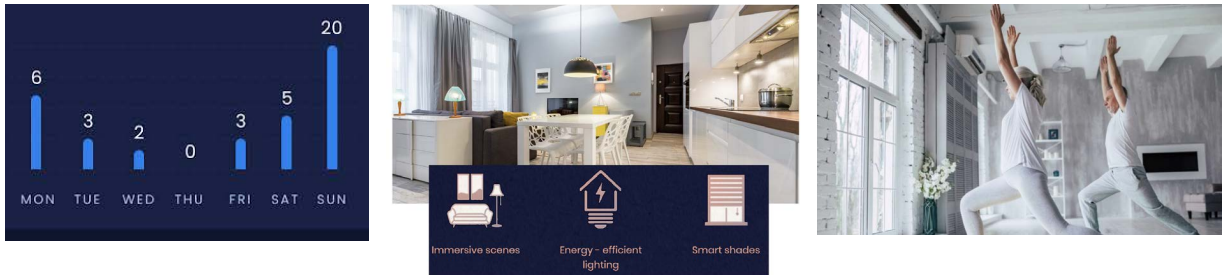

*Figure 16. Coach for an Active Lifestyle: AI-assistants provide activity reports and set the home environment to help residents sleep well, exercise, and wake up to natural light.*

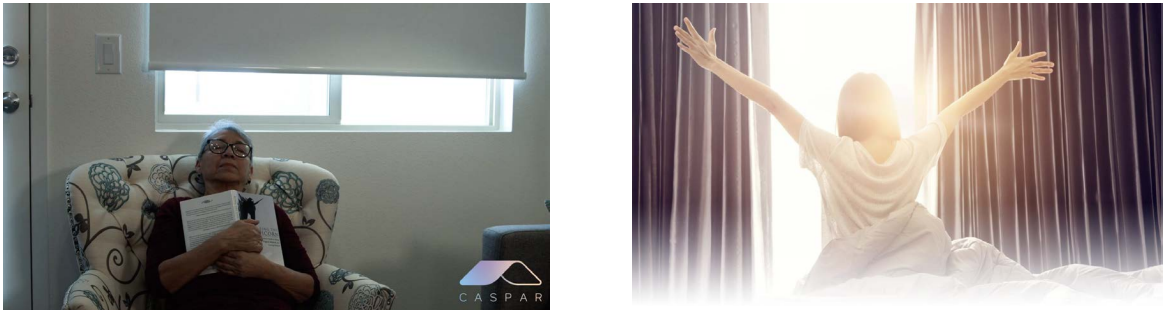

*"It was a Godsend to have Caspar when I needed it most,"*

*- Kate, an 88-year old resident, who was bed-ridden with an illness.*

*Figure 17. AI-enabled pervasive automation helps manage home's climate, lighting and shades. When seniors are ill, even basic tasks like turning on a light become a challenge.*

## 4.2 Maintenance, Services and Visitors

At Revel, we enable proactive maintenance which reduces unnecessary apartment visits to minimize physical contact. For example, if a bulb or thermostat breaks or if there is water leak, an automated ticket is issued to the maintenance staff; thus a repair visit can be scheduled to minimize contact between resident and staff. The resident is alerted when the staff visits through our awareness system (Section 4.3). As staff enters, the home automatically adjusts the environment avoiding maintenance staff touching switches or thermostats – thus reducing potential infection spread.

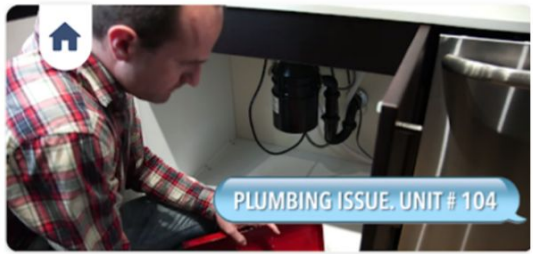
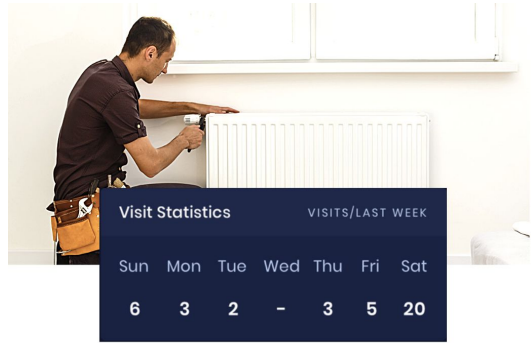

*Figure 18. Proactive maintenance with alerts and tracking. The resident knows about visitors, and for the onsite manager it gives visibility into the staff's efficiency and can lead to increased safety.*

Our outdoor awareness cameras send a notification to the residents when visitors, cleaning staff, or packages are at their door. Using AI face recognition, residents can differentiate between known and unknown visitors. Additionally residents that are ill or self isolating are informed when a package or food delivery arrives without the resident needing to come into contact with the delivery person.

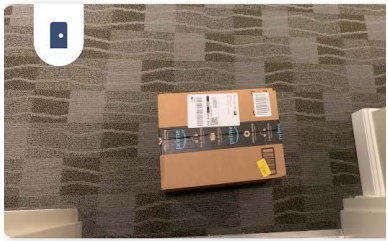
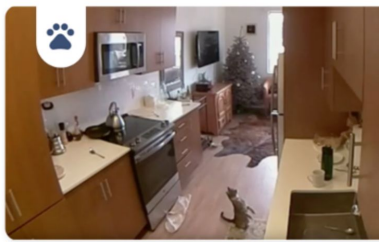
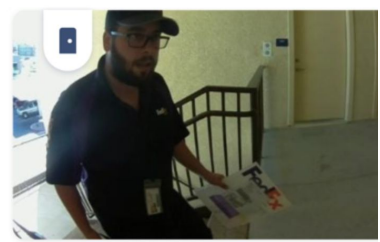

*Figure 19. Notification of visitors and deliveries. Residents know when their food or package delivery arrives and can choose when to bring it in, thus minimizing contact.*

## 4.3 Common Areas, Self-guided Tours and Remote management

As discussed in Section 2.3, common areas in senior living communities play a crucial role – allowing residents to interact with each other. Revel holds events in a number of common areas, such as art classes in the activity room, game nights in the dining room, or yoga classes in the studio. Unfortunately, during the Covid-19 pandemic, common areas were the primary source of infection and therefore had to be closed.

Caspar AI equips common areas with smart sensors and AI-driven analytics. Currently, our system provides remote management of these common areas, with hands-free automation that sets the desired ambiance – optimal settings for Yoga classes or Gym settings for exercise. This allows staff to manage areas safely and remotely, increasing engagement and satisfaction amongst residents. We have not deployed a comprehensive solution to address the full spectrum of pandemic concerns yet – we do not foresee any obstacles to having an end-to-end pandemic solution this year.

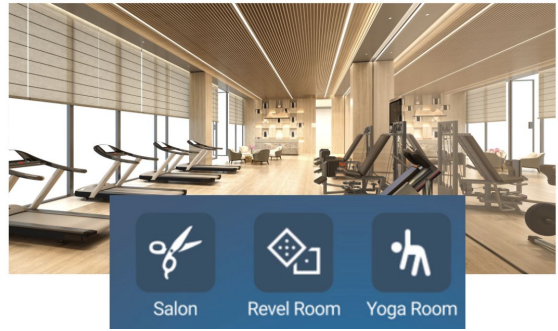

"We allow only a limited number of visitors in the activity rooms. With Caspar AI, I do not have to go room to room to adjust temperature and ambiance."

- *an onsite staff member at Revel.*

*Figure 20. At Revel, common areas are managed remotely with AI-enabled automation.*

In addition to common area management, our automation provides self-guided tours that allow visitors to experience model units without staff members present. This increases the safety of visitors and staff members. Management can set the desired ambiance for model units across multiple properties with a single button.. Once visitors leave, the apartment automatically switches devices off to save energy. Finally, the amount of energy saved is reported to help management make the operations sustainable.

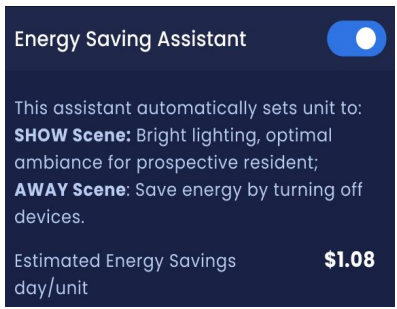
*A single toggle puts the building into energy-saving mode, while maintaining the desired functionality.*

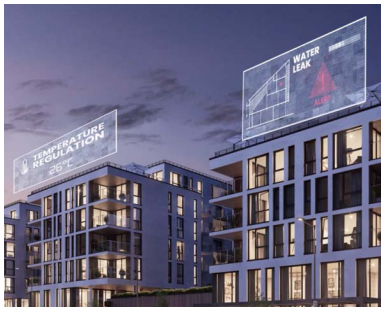
*Remote management: For common areas and vacant units, you have an awareness of incidents.*

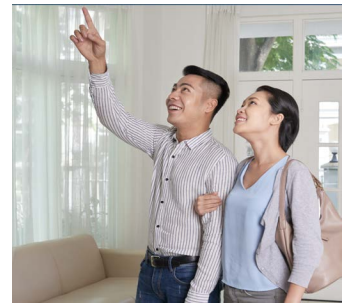
*Self-guided tours: An optimal ambiance is set automatically when prospective residents are touring units.*

*Figure 21. At Revel, the common areas and vacant units at Revel get easier to manage. Automation enables remote building management, while saving energy costs.*

# 5 Savings and Value Add

In this section, we discuss the savings for senior home operators and the value added for residents with pervasive automation. With pervasive automation, you can get a value of 10% of the monthly rent.

As an illustration, let us consider the senior living community described in Section 4 with the Caspar AI system, where the monthly rent is about $2,800. We estimate a net value of $284 per month per apartment as described below. (In our studies at other properties, we have seen that this net value varies proportionally with the rent.)

*Table 1. Pervasive automation can achieve savings and value add of about 10% of monthly rent.*

| | |
|---|---|
| Savings for caregiving staff and operators | $174 |
| Value to the residents for safety and refined living services | $145 |
| Costs: customer support, 5-year amortized hardware | -$35 |
| **Net monthly value per apartment** | **$284** |

The savings come primarily from the man-hours saved in automating the existing manual services (Table 2 that assumes a rate of $20/hr). In some cases automation even provides a superior offering, e.g., 24/7 activity notification for the resident's family vs daily manual check-in. In addition to the man-hours saved, there are additional benefits because a safe apartment minimizes legal risks, reduces insurance costs, and helps rent the apartment faster.

*Table 2. Estimated monthly savings for a 100-unit senior living property.*

| Feature | Description | Man-hours | Savings |
|---|---|---|---|
| Automated Check-in | Each check-in takes a minimum of 7 minutes per resident. Automated check-in saves time with a 24/7 sensing. | 350 | $7,000 |
| Carecircle notifications | Carecircle receives notifications automatically, saving 5 hours daily for staff on calls with the resident's family. | 150 | $3,000 |
| Maintenance Alerts | We estimate saving of about 1 hour per month per unit for proactive maintenance and diagnosing maintenance issues. | 100 | $2,000 |
| Remote Building Management | Video clips give visibility about the visitors and property events. Also remotely manage the vacant units. | 100 | $2,000 |
| Common Areas | Automation saves about 2 man-hours per day in setting the optimal environment in common areas and scheduling. | 50 | $1,000 |
| Remote Auto-show | Visitors can visit units themselves. The staff can focus on key sales points while being safe. | 100 | $2,000 |
| Energy Savings | For the 5% of units that are not occupied, about $38 per month each is wasted. Further savings in common areas. | - | $400 |
| Faster renting | A differentiated apartment that addresses safety concerns and provides easy-to-use amenities rents faster | - | - |
| Risks and Insurance | There is additional value in reducing risks, including legal risks due to infection, insurance risks, or water leaks. | - | - |
| Total monthly savings per month on a 100-unit building | | | $17,400 |
| **Total monthly savings per month per apartment** | | | **$174** |

Additionally, residents also get a variety of features for comfortable, safe and engaged living. The table below lists the monthly fees that residents pay on an average for these services. For some features, including Carecircle awareness, residents have told us they would pay up to $200 per month, but we have used a more conservative number of $50. When residents have reduced mobility, they also have a larger need for automated home control and AI-enabled assistants – the value they place is often higher than $50, but we have used a more conservative estimate of $25.

*Table 3. Monthly value added for residents per apartment.*

| Feature | Description | Value |
|---|---|---|
| Carecircle | 24/7 awareness about the resident's safety | $50 |
| Control and Assistants | Voice- and app-based control for climate, lighting, shades, etc. Assistants for sleep, wake-up, exercise, cooking, etc. | $25 |
| Events and video clips | Video clips for visitors, packages, and pet activities for security. Ability to see pets and extend to a pet feeder. | $50 |
| Engaged Living | Intelligent Assistants for engaged living. Adapt the home environment, e.g., open curtains for natural light in the morning. | $20 |
| **Total monthly value add to resident per apartment** | | **$145** |

As we enhance our services and add more features from our vision for the future, which was presented in Section 3, we have an opportunity to achieve even greater value for both operators and residents in the form of savings and advanced AI-enabled features.

# 6 Conclusion

Pervasive automation allows you to design and operate senior living facilities in a dramatically different way. The conventional approach of high-touch caregiving is no longer viable after Covid-19, so steps must be taken to fundamentally alter how senior living facilities operate.

Pervasive automation can take your facilities from **high-touch** to **no-touch**. It can enhance the safety and efficiency of senior living staff – most services in homes and common areas can be managed remotely. Residents' carecircle can receive automated alerts about incidents such as inactivity, falling, and the stove being left on. Automation helps with daily tasks such as turning off the lights at night and provides data that can help signal early signs of diseases such as dementia. Residents can live a healthy and engaged lifestyle with improved sleep quality and helpful information about their level of activity. The home motivates them to maintain an active lifestyle, e.g., by opening the curtains in the morning to set the mood with natural light.

This is not an abstract vision. We currently provide senior living facilities, with automation enabled services such as automated alerts, daily check-in and common area management. This has increased the safety of the occupants and has reduced operating costs. With their homes and common areas automated, the residents live a healthy and engaged lifestyle without requiring unnecessary physical contact.

With pervasive automation you can mitigate the risks that you, your staff, and your residents face and be on the forefront of this evolutionary shift in the senior housing market.